\documentclass[structabstract]{aa}  
\usepackage{graphicx}
\usepackage{txfonts}
\usepackage{natbib}
\usepackage{textcomp}
\bibpunct{(}{)}{;}{a}{}{,}

\begin{document}
\title{The blue UV slopes of z$\sim$4 Lyman break galaxies: implications for the corrected star formation rate density}

   \author{M. Castellano \inst{1}
   \and
   A. Fontana \inst{1}
   \and
   A. Grazian \inst{1}
   \and
   L. Pentericci \inst{1}
   \and
   P. Santini \inst{1}
   \and
   A. Koekemoer \inst{2}
   \and
   S. Cristiani  \inst{3}
   \and
   A. Galametz \inst{1}
   \and
   S. Gallerani \inst{1}
   \and
   E. Vanzella  \inst{3}
    \and
   K. Boutsia \inst{1}   
   \and
   S. Gallozzi \inst{1}
   \and
   E. Giallongo \inst{1}
   \and
   R. Maiolino \inst{1}
   \and
   N. Menci \inst{1}
   \and
   D. Paris \inst{1}
          }

   \offprints{M. Castellano, \email{marco.castellano@oa-roma.inaf.it}}

\institute{INAF - Osservatorio Astronomico di Roma, Via Frascati 33,
I--00040, Monteporzio, Italy \and Space Telescope Science Institute, 3700 San Martin Drive, Baltimore, MD 21218 \and INAF - Osservatorio Astronomico di Trieste, Via G.B.
Tiepolo 11, 34131 Trieste, Italy.
}

   \date{Received... accepted ...}

   \authorrunning{Castellano et al.}
   \titlerunning{The blue UV slopes of z$\sim$4 Lyman break galaxies}

 
  \abstract
{The study of the dust extinction in high-redshift galaxies is fundamental to obtain an estimate of the corrected Star Formation Rate Density (SFRD) and to put constraints on galaxy evolution models.} 
   {We plan to analyse dust extinction in Lyman Break Galaxies (LBGs) by introducing a new and more reliable approach to their selection and to the characterization of their distribution of UV slopes $\beta$, using deep IR images from HST. We fully discuss the methodology and the results focusing on a robust sample of z$\sim$4 LBGs.}
   {We exploit deep WFC3 IR observations of the ERS and HUDF fields over GOODS-South, combined with HST-ACS optical data, to select z$\sim$4 LBGs through a new (B-V) vs. (V-H) colour diagram. The UV slope of the selected galaxies is robustly determined by a linear fit over their observed I, Z, Y, J magnitudes, coherently with the original definition of $\beta$. The same fit is used to determine their rest-frame UV magnitudes $M_{1600}$ through a simple interpolation. We estimate the effect of observational uncertainties with detailed simulations that we also exploit, under a parametric maximum-likelihood approach, to constrain the probability density function of UV slopes PDF($\beta$) as a function of rest-frame magnitude.
   }
   {We find 142 and 25 robust LBGs in the ERS and HUDF fields respectively, limiting our sample to S/N(H)$>$10 objects.  Our newly defined criteria improve the selection of $z\sim$4 LBGs and allow us to exclude red interlopers at lower redshift, especially z$\sim$3-3.5 objects. We show that the use of a linear fit to estimate $\beta$ and an accurate characterization of observational effects are required in such analysis of flux limited samples. We find that $z\sim$4 LBGs are characterized by blue UV slopes, suggesting a low dust extinction: all $L<L^*$ galaxies have an average UV slope $\langle\beta\rangle\simeq -2.1$, while brighter objects only are slightly redder ($\langle\beta\rangle\simeq -1.9$). We find an intrinsic dispersion $\simeq 0.3$ for PDF($\beta$) at all magnitudes. The SFRD at z$\sim$4 corrected according to these estimates turns out to be lower than previously found: log(SFRD)$\simeq-1.09~M_{\odot}/yr/Mpc^3$. Finally, we discuss how the UV slope of z$\sim$4 galaxies changes as a function of the dust-corrected UV magnitude (i.e. SFR). We show that most galaxies with a high  SFR ($\gtrsim80~M_{\odot}/yr$) are highly extincted objects. Among galaxies with lower SFR, we detect many with a much lower amount of reddening, although  current observational limits prevent us from detecting those with high  extinction, if they exist.
   }{}


   \maketitle
\section{Introduction}
It is now nearly 20 years since the first surveys explicitly dedicated to detect galaxies at $z\simeq 3$ have been completed \citep{Steidel1992,Steidel1993,Giavalisco1994,Steidel1995}.  In the early 90's, these seminal works have introduced the so-called ``Lyman Break'' method to select rest--frame UV bright objects, i.e. star--forming, moderately absorbed galaxies, at $z\simeq 3$ and at higher redshifts.  
In the following years, this technique has been widely applied by surveys covering either progressively wider areas \citep[e.g.][]{Steidel2003,Ouchi2004,McLure2009} to enlarge the volume sampled, or reaching fainter magnitudes \citep[e.g.][]{Bouwens2007,Reddy2009} to detect the faintest galaxies, or extending to longer wavelengths to explore the highest redshifts \citep[e.g.][]{Bouwens2004,Castellano2010,Bouwens2011}. In the meantime, spectroscopic follow-up campaigns of Lyman Break Galaxies (LBGs) samples have shown that this technique is extremely efficient and reliable up to the highest redshifts \citep[e.g.][]{Steidel1996,Vanzella2009,Stark2010,Vanzella2011,Pentericci2011}.

Some of the primary goals of these surveys has been the statistical analysis of high redshift galaxies, in particular of their UV Luminosity Function (LF) from $z\simeq$3 to $z\simeq$7 and beyond \citep[][]{Steidel1999,Bouwens2007,McLure2009,Bunker2010,Grazian2011}, and their contribution to the Star Formation Rate Density (SFRD). 

Understanding the physical properties of UV--selected galaxies is also fundamental to derive the global properties of the Universe at these high redshifts. In particular, to estimate their contribution to the total SFRD one needs to apply the proper conversion between the observed UV rest frame luminosity and the ongoing Star Formation Rate (SFR). This conversion factor depends on the physical properties of the stellar population (IMF, metallicities and ages) and on the amount of dust extinction. 
The slope of the UV continuum in LBGs provides precious information to measure these quantities.  Apart from weak absorption lines and other minor features, the spectrum of a star--forming galaxy longward of the Ly$\alpha$ wavelength can be reasonably approximated by a power law until 3200\AA~for most combinations of stellar properties and attenuation curves \citep{Calzetti1994,Meurer1999,Calzetti2000,Gallerani2010}. The actual slope is the result of several factors involved in the stellar mixture in the galaxy: the total amount and composition of dust grains, the metallicity $Z$, the initial mass function (IMF), and the star formation history (SFH).  Lower metallicity stars, top heavy IMFs and young ages all tend to produce bluer spectra than higher metallicities, standard IMFs and aging stellar populations. Nebular continuum emission can also contribute to redden the slope for galaxies. We note that nebular emission is expected only if the escape fraction of UV ionizing photons is small, as currently observed in most of the galaxies at $z\leq 3.5$ \citep{Vanzella2010,Boutsia2011}.  However, dust absorption is likely the strongest reddening factor, and the observed slope is often directly converted into dust extinction assuming standard stellar populations. 

In all cases, the UV spectrum is well approximated by a simple power law $f_{\lambda}= \lambda^\beta$: the accurate determination of the exponent $\beta$ is the central goal of the present paper. For reference, a dust-free stellar population with solar metallicity and constant star-formation rate has $\beta\simeq -2.2$.  Clearly, it is impossible to constrain all the physical quantities mentioned above from a single parameter. Nevertheless, a robust measure of $\beta$ is the first mandatory step to constrain these free parameters and derive the global properties of LBGs.

Despite the large samples collected in the last years, only a few estimates of the statistical distribution of $\beta$ are available today, and no firm conclusion has been reached on its evolution with luminosity and redshift. 
Previous determinations have been provided by \citet{Bouwens2009a}, that found a strong decrease of the average UV slope both with rest-frame magnitude and redshift from $z\sim2$ to $z\sim6$, and by \citet{Bouwens2010b} presenting evidence for very blue UV slopes in their z$\sim$7 sample. At z$\sim$6-8 \citet{Finkelstein2010} found blue UV slopes consistent with dust-free normal stellar populations. On the other hand, \citet{Dunlop2011} found that UV slopes are consistent with $\beta\simeq-2$ at $z\sim5-7$, and argue that very steep values found at high redshift are due to the effects of photometric scatter and uncertainties. \citet{Wilkins2011} reached a similar conclusion from their $z\sim$7 sample, but at z$\sim5$ found average slopes redder than both \citet{Dunlop2011} and \citet{Bouwens2009a}. 
At $z\sim4$ \citet{Ouchi2004} found an average E(B-V)$\simeq0.15$ and no significant dependence on UV magnitude for $L>L_*$ LBGs, while \citet{Lee2011} found a decreasing extinction, from E(B-V)$\simeq$0.22 at $L\approx 4-5 L_*$ to E(B-V)$\simeq$0.07 at $L\approx L_*$.

Discrepancies among different works are probably due to the known limits of LBG selection techniques \citep[e.g.][]{Stanway2008} and to uncertainties in the measure of the UV slope, often based on a single colour. 
These limitations can be overcome by assembling deep near IR observations of large LBG samples: these are required to strengthen colour selections and to fully sample the UV continuum up to $\simeq 3000$\AA~providing a proper estimate of $\beta$. For this reason, early results based on optically selected samples are intrinsically prone to larger systematic effects. Even more importantly, however, they suffer of a conceptual limitation. For a given dereddened luminosity (or equivalently a given SFR), even a modest dust absorption produce a significant dimming of the galaxy, which may escape detection in optically selected surveys.

In this paper we take advantage of the new IR observations  carried out with WFC3  on board of HST. We show that these deep observations in the Y, J and H bands provide significant advantages over previous ones in three different aspects:

$\bullet$ They make possible a much cleaner selection of LBGs, compared to the traditional criteria that use optical bands only. This allows us to reject lower redshift interlopers with much higher reliability.

$\bullet$ They yield a robust determination of the UV slope, providing both a large leverage in wavelength and a sizeable number of independent determinations of the flux  (of each object) at different wavelengths.

$\bullet$ They allow us to detect galaxies with a larger slope for a given dereddened luminosity. In this respect,  the H--band that we use to detect galaxies constitutes the optimal choice since it corresponds to the reddest portion of the UV rest-frame domain ($3200$\AA~rest--frame at $z\simeq 4$), significantly redder than the $1500-1800$\AA~of optically selected samples.  Hence our samples are intrinsically more complete in terms of content of reddened galaxies than optically selected ones, although this is currently limited by the current depth of optical observations used for the colour selection.

To fully exploit these potential advantages, we apply a tailored colour selection criterion to our H--band selected catalogue, and we measure $\beta$ with a straightforward linear fit to the observed magnitudes. In principle, similar results could be derived using a SED fitting approach on the same H--selected sample \citep[e.g.][]{Fontana2003}, where the slope depends on the free parameters mentioned above: IMF, metallicity, age, E(B-V). We decide to adopt a simpler method for two reasons. First, as mentioned above, the four free parameters are to a large extent degenerate, even using additional constraints from the Spitzer magnitudes. Our approach determines a single parameter ($\beta$) from four observed magnitudes. Further physical interpretation can be derived from the observed values of $\beta$ alone. In addition, a complete analysis requires extensive Monte Carlo simulations, that are required to take into account the involved systematic effects. Selecting galaxies with a simple colour selection criterion and adopting a single free parameter makes these simulations affordable.

In this paper we apply our method to $z\simeq 4$ LBGs. In future papers, we plan to extend the analysis to higher redshifts, using also the forthcoming data sets provided by the CANDELS survey \citep{Grogin2011,Koekemoer2011}.

The plan of the paper is the following: in Sect.~\ref{data} we present our data-set from which we select a sample of B-dropout galaxies applying a newly defined IR-based colour selection, discussed in Sect.~\ref{sample}. In Sect.~\ref{slopes}, we present the estimate of UV slopes and rest-frame magnitudes for our objects, along with the simulations used to evaluate systematic effects and uncertainties. These simulations are exploited in Sect.~\ref{analysis} to constrain the distribution of $\beta$ at different UV magnitudes, from which a dust-corrected value of the SFRD is estimated in a straightforward way (Sect.~\ref{sfrd}). In Sect.~\ref{UVcorr}, we discuss how the UV slope of z$\sim$4 galaxies in our sample changes as a function of their dereddened UV emission. Finally, a summary is given in Sect.~\ref{summary}.   

Throughout the whole paper, we adopt a
\citet{Calzetti2000} extinction law, unless otherwise specified. Observed and rest--frame magnitudes are in
the AB system, and we adopt the $\Lambda$-CDM concordance model
($H_0=70km/s/Mpc$, $\Omega_M=0.3$, and $\Omega_{\Lambda}=0.7$).

\section{Data and Photometry}\label{data}
\begin{figure*}[!ht]
   \centering
   \includegraphics[width=17cm]{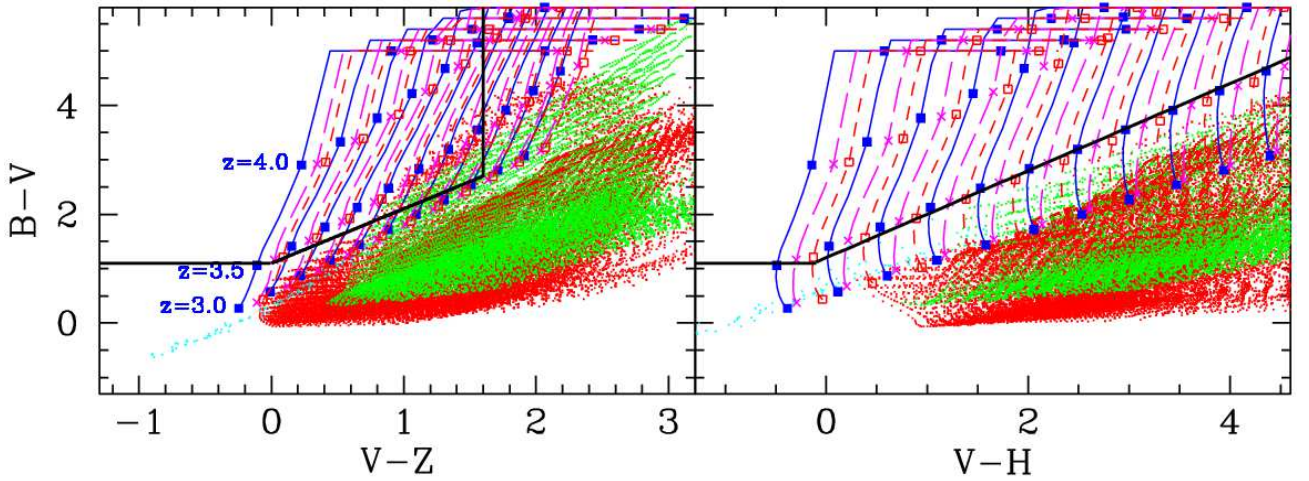}
   \caption{The (B-V) vs (V-H) colour selection window adopted in the present paper (right) along with the commonly adopted one based on the (V-Z) colour (left). Blue filled squares and lines are for a CB07 LBG model with $Z/Z_{\odot}=0.2$ and Age=100Myr at increasing redshift and E(B-V) respectively. Models having solar metallicity and Age=100Myr and Age=300Myr are represented in magenta (crosses and long-dashed lines), and red (empty squares and short-dashed lines), respectively. Red and green points indicate lower redshift passively evolving and dusty star-forming galaxies respectively. See main body for details.}
         \label{fig1}
\end{figure*}
In the present paper we exploit data obtained in the WFC3/IR ERS2 program \citep{Windhorst2011} and the ACS and WFC3/IR UDF programs \citep{Beckwith2006,Bouwens2010c}, where the data were processed and combined as described in \citet{Koekemoer2011}, along with the full public ACS data available in the GOODS-South field \citep{Giavalisco2004}.
Full details on image reduction and catalogue building are given in
\citet{Grazian2011} and \citet{Santini2011}: we summarize here the main properties of our dataset.

The GOODS-ERS WFC3/IR dataset (HST Program ID 11359) covers $\sim 40 sq. arcmin.$ in the $Y_{098}$, $J_{125}$, $H_{160}$ filters. 
It is made of 10 contiguous pointings in the GOODS-South field observed during 60 HST orbits, for a total of 4800-5400s per pointing and filter. The astrometry of WFC3 ERS images was
bound to the GOODS-S v2.0 z band \citep{Giavalisco2004}.
The limiting magnitudes are Y=27.3, J=27.4, and H=27.4 at $5\sigma$ in an area of
$\sim 0.11$ $arcsec^2$ (corresponding to 2FWHM).

The HUDF WFC3/IR dataset (HST Program ID 11563) is a total of 60 HST
orbits in a single pointing of 4.7 $sq. arcmin.$ \citep[e.g.][]{Oesch2009b,Bouwens2009b}
in three broad-band filters
(16 orbits in $Y_{105}$, 16 in $J_{125}$, and 28 in $H_{160}$). 
The HUDF images were reduced in the same way as the ERS ones and they have been
astrometrically matched using the ACS HUDF sources.
For the $Y_{105}$ filter, we have selected a subset of 7 visits out of 9
that were not badly affected by persistence from bright sources.
The magnitude limits of HUDF WFC3 observations are Y=29.0, J=29.2, and
H=29.2 total magnitudes for point like sources at 5$\sigma$
in an aperture of $\sim 0.11$ $arcsec^2$.

In both fields we use the available BViz ACS public data: the latest V2.0 version of ACS BViz images of GOODS-South released by the
STSci (M. Giavalisco and the GOODS Team, in preparation) for the ERS, and the deep BViz ACS data presented in \citet{Beckwith2006} for HUDF.
To ensure precise colours, the 
ACS bands were smoothed with an appropriate kernel to match
the resolution of the WFC3 images.

We perform object detection on the H$_{160}$ band, obtaining  10946 and 2418 sources for the ERS and HUDF, respectively.
Their total magnitudes have been computed using \verb|MAG_BEST| of
SExtractor \citep{Bertin1996} for galaxies more extended than 0.11 $arcsec^2$, while
circular aperture photometry (diameter equal to 2 times the WFC3
FWHM) is used for smaller sources. The magnitudes which are computed
in a circular aperture have been corrected to total by
applying an aperture correction of 0.4 magnitudes. Colours in BVIZYJ have
been measured running SExtractor in dual image mode, using isophotal
magnitudes for all the galaxies.

\section{A new approach to the selection of z$\sim$4 LBGs}\label{sample}

We define the sample of z$\sim$4 galaxies by applying the Lyman-break technique. The main spectral feature that enables
the identification of high-z galaxies is the drop shortward of the Lyman-$\alpha$, which, at z$\gtrsim$3.5, is redshifted at $\lambda\gtrsim 5500$\AA: the Lyman break can thus be effectively sampled by the B and V bands (B-dropout galaxies). The usual colour-selection of z$\sim$4 LBGs is carried out on a (B-V) versus (V-Z) diagram \citep[e.g.][]{Giavalisco2004,Bouwens2009a,Vanzella2009} to separate LBGs from lower redshift Balmer break and dusty galaxies which show redder colour at longer wavelengths. However, as already noted by \citet{Beckwith2006} the redshift selection range provided by these criteria has a non-negligible dependence on the amount of extinction.  For example, \citet{Beckwith2006} (Tab. 8 and 9) show that a 100Myr star-forming model with E(B-V)=0.15 can be included in the selection from lower redshifts ($z_{min}=3.3$) than a dust-free one ($z_{min}=3.5$).

\subsection{Colour selection}\label{selection}
In this paper, thanks to the very deep WFC3 near infrared images, we adopt a modified version of the B-dropout selection which is based on the (V-H) colour instead of the (V-Z) one. We used models taken from Charlot and Bruzual 2007 \cite[][hereafter CB07]{Bruzual2007a,Bruzual2007b} to empirically define colour criteria allowing for a clean selection of $z\sim4$ LBGs :
\begin{eqnarray*}
(B-V) &>& 1.1 \\
(B-V) &>& 1.2+0.8\cdot(V-H) .
\end{eqnarray*}

\begin{figure}[!ht]
   \centering
   \includegraphics[width=9cm]{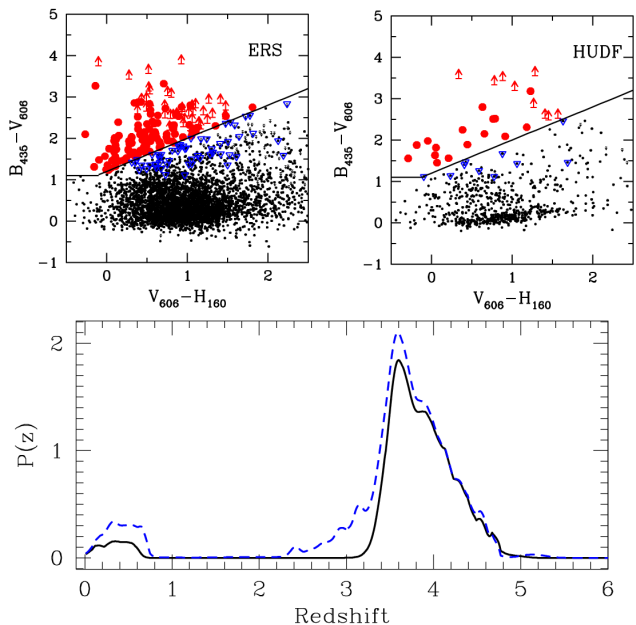}
   \caption{\textit{Upper panels:} colour-colour diagrams for B-dropouts in the ERS (left) and HUDF (right) fields. Selected objects are marked in red, the blue points indicate B-dropout candidates in the (B-V) vs (V-Z) selection window excluded by the present analysis. \textit{Lower panel:} combined photometric redshift probability distribution of our ERS B-dropout candidates (black line) and of galaxies selected from the (B-V) vs (V-Z) diagram (blue dashed line): the (V-H) selected objects have a lower probability of having $z_{phot}\lesssim3.5$ solutions.}
         \label{fig2}
\end{figure}

In Fig.~\ref{fig1} we compare the commonly adopted selection window (left panel) to the one used here (right panel) for the CB07 models. The blue filled squares indicate the position of a reference LBG model with $Z/Z_{\odot}=0.2$ and Age=100Myr from z=3 to z=5 at 0.5 intervals, with the blue continuous lines, from left to right, showing the effect of an increasing extinction from E(B-V)=0.0 to E(B-V)=1.0 at 0.1 intervals. Magenta and red tracks are for LBG models with $Z/Z_{\odot}=1.0$ and Age=100Myr and 300Myr respectively.
Red points indicate galaxies at z=0.0-3.5 with short
star formation exponential timescales ($0.1-1$ Gyrs) and ages $>1$ Gyr, while green symbols indicate constant star-forming
models with $0.5<E(B-V)<1.5$ in the same redshift range.
It appears evident that the (V-Z) colour cannot efficiently discriminate between B-dropout and lower redshift contaminants, since z$\gtrsim3.5$ LBGs with moderate E(B-V) and both lower redshift dusty star-forming galaxies and passive ones occupy the same region of the colour diagram. This effect potentially leads to a contamination in the sample of the more extincted B-dropout galaxies, biasing the estimate of the UV slope distribution at z$\sim$4 and cannot be completely avoided by redefining the relevant color criteria. The (V-H)-based criterion instead allows us to remove the reddest potential interlopers from the sample. 

Moreover, the (V-H) colour provides a cleaner LBG selection window: a careful look at the reference LBG tracks in Fig.~\ref{fig1} shows that the (B-V)-(V-H) criteria are, in principle, unbiased since they are met by LBGs at z$>3.5$ regardless of their E(B-V), age and metallicity.  Specifically, there is no need to adopt \textit{a priori} a cut in the (V-H) colour, contrary to the (V-Z) one, and any intrinsic constraint on the (V-H) colour due to observational limits corresponds to the same cut in E(B-V) for galaxies of different redshifts. On the other hand, the usual (B-V)-(V-Z) cut is redshift-dependent, with galaxies at increasing E(B-V) meeting the selection criteria at decreasing redshifts, in agreement with the findings of \citet{Beckwith2006}.

\subsection{The ERS+HUDF B-dropout sample}\label{samplesubsect}
Our LBG selection is naturally limited by the capability of accurately measuring large (V-H) colour terms, that is by the relative depth of the optical and IR data-set. We limit our sample to S/N(H)$>$10$\sigma$ to consider only objects with accurately measured colours: this choice implies that the V-band magnitude of an object with E(B-V)=0.5 (corresponding to (V-H)$\sim$2, i.e. $\beta \sim$0) is measured at $\sim$3$\sigma$. On the other hand, to exclude $z\gtrsim4.5$ galaxies, we do not consider objects with S/N$<$3 in the V band: in practice, given the conservative S/N cut in the H band, the latter criterion only removes one bright V-dropout galaxy from the ERS selection. Finally, to ensure that no AGN are present among the selected objects, we remove three point-like sources (Sextractor \verb|CLASS_STAR|=0.99) from the ERS sample and check on the 4Ms CDFS images that none of our candidates present X-ray emission (Fiore et al. 2011, subm.). The final sample consists of 142 objects at  $23.9<H<26.5$ from the ERS field, and 25 objects (24.8$<H<$27.4) from the HUDF. 

We show in Fig.~\ref{fig2} the relevant colour-colour diagrams and the differences between the criteria adopted in the present paper and the usual B-dropout selection: our selected candidates are indicated by red points, while the blue symbols mark the position in the diagram of objects that would be included by a standard (V-Z) colour-selection, but are excluded in our sample. 
Following the previous analysis of the performance of the two different selections on CB07 models, the position of the latter in the diagrams is consistent with them being either low redshift interlopers or U-dropouts at $z\sim3.0-3.5$. 

This can be verified through a simple test by exploiting our ERS photometric redshift catalogue that also exploits deep K band and IRAC observations \citep{Santini2011}. We note that the photometric redshifts in the ERS field have been obtained through the well tested $\chi^2$ minimization procedure outlined in \citet{Grazian2006} and \citet{Santini2009}, which is based on a different set of synthetic spectral templates \citep[PEGASE 2.0 by ][]{Fioc1997} than those (CB07) used in Fig.\ref{fig1}.  We determine the combined photometric redshift distributions P(z) for the two selection criteria as the sum of the redshift probability distribution of the relevant B-dropout candidates. As shown in the lower panel of Fig~\ref{fig2}, when objects selected through the (V-Z) colour are considered (blue dashed curve), P(z) displays a lower redshift tail extending out to $z\sim2.5$.  On the other hand, the (V-H) selected sample (black continuous line) is characterized by a sharper P(z) at high redshift and has a lower probability for secondary solutions at $z<1$: by integrating P(z) we estimate that the contamination in a sample selected through the (V-Z) colour is more than twice than in a (V-H) selected one, with 32\% objects at $z<3.5$ (14\% at $z<3.0$), compared to 15\% (5\% at $z<3.0$) in the latter case.  Photometric redshifts solutions at $z<1$ are due to the similarity between the optical and near-IR SEDs of Balmer break galaxies at these redshifts, and those of B-dropout galaxies: in this respect, an improved colour-based selection is not feasible, since it would require deeper Spitzer data to enable a thorough study of the LBG population.

We also exploit ERS photometric redshifts to explore different criteria for both the (B-V)-(V-H) diagram and the (B-V)-(V-Z) one. In particular, we consider 1) 172 candidates selected through a less conservative (V-H) criterion, i.e. having the redder selection boundary $(B-V)>1.1+0.7\cdot(V-H)$, and 2) 129 B-dropouts isolated through a ``steeper'' (V-Z) colour cut aimed at avoiding a redshift-dependent selection of galaxies with different E(B-V)\footnote[1]{defined on the basis of CB07 LBG tracks: (B-V) $>$ 1.1 $\wedge$
(B-V) $>$ 1.1+1.5$\cdot$(V-Z) $\wedge$ (V-Z) $<$ 1.6}. In the first case we find an increase of the contamination from z$\sim3-3.5$ LBGs (P(z$<$3.5)=18\%), indicating that this new boundary is indeed extending towards lower values the redshift selection window, consistently with the LBG tracks displayed in Fig.~\ref{fig1}.  In the second case, we still find a larger contamination than in our original (B-V)-(V-H) sample, with a 26\%(12\%) probability for objects at $z<3.5$ ($z<3.0$). These findings are in agreement with the tests on CB07 models (Sect.~\ref{selection}) and show that our proposed criteria provide a more reliable selection for B-dropout galaxies.

Objects having spectroscopically confirmed redshift further validates the efficiency of our method. Out of ten objects with $3.5 \leq z_{spec}\leq 4.5$ and S/N(H)$>$10, the (V-H) selected sample includes nine with $3.58 \leq z_{spec}\leq 4.4$ and only fails to select a LBG at $z_{spec}=4.41$. Instead, the (V-Z) colour criterion also includes a galaxy at $z_{spec}=3.11$ along with those already present in our sample.
This is also consistent with the result of intensive follow-up spectrocopic campaigns of B-dropout samples across all the GOODS-South field:  \citet{Vanzella2009} finds that 8 out of 46 B-dropout galaxies initially selected through the (B-V)-(V-Z) diagram lie in the range $3.3\leq z_{spec} \leq 3.5$. A similar result is evident from Fig. 7 of \citet{Stark2010} showing the redshift distribution of their spectroscopic sample over GOODS-North. 
We show in the next section and in Fig.~\ref{fig6} how the inclusion of these potential interlopers in our sample would affect the determination of the UV slope distribution.

\section{Estimating UV slopes and rest-frame magnitudes}\label{slopes}
\begin{figure}[!ht]
   \centering
  \includegraphics[width=9.5cm]{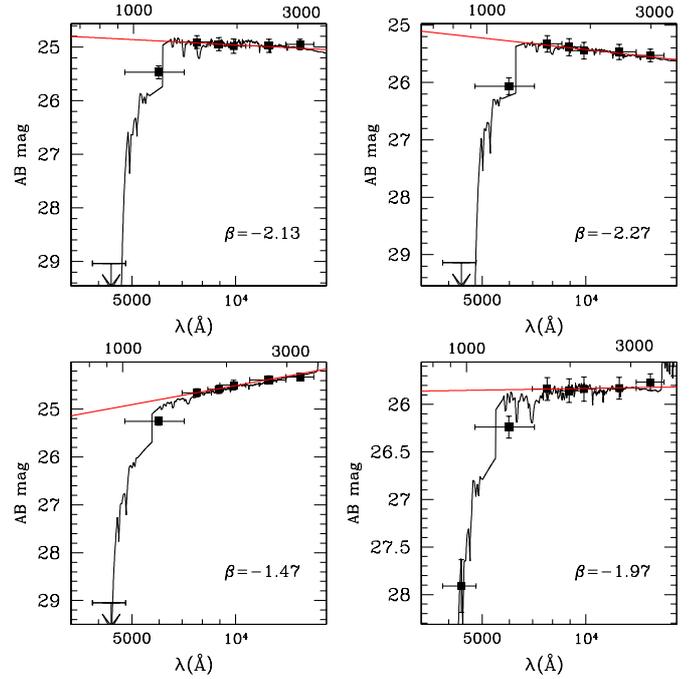}
   \caption{A comparison of the fitted UV slope (red continuous line) and best-fit SED of four objects in our sample having best-fit photometric redshift in the range $z_{phot}=3.5-4.1$. The observed magnitudes (filled squares and errorbars) are, from left to right: B, V, I, Z, Y, J, H. The linear-fit is performed on I, Z, Y, J. The lower and upper horizontal axis show the observer-frame and the rest-frame wavelength, respectively.}
         \label{fig3}
\end{figure}

\begin{figure}[!ht]
   \centering
\includegraphics[width=8cm]{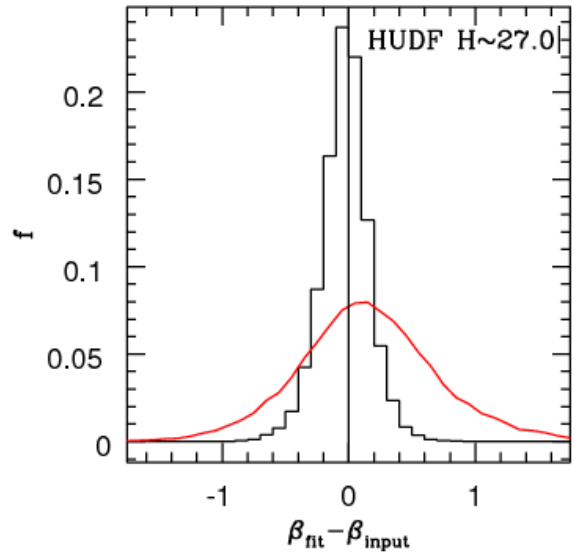}
   \caption{Comparison of uncertainties on $\beta$ derived from a linear fit to the observed I, Z, Y, J magnitudes (histogram) and from the (I-Z) colour only (red curve) for simulated objects with observed UV slope $\beta_{fit}$. The real UV slope $\beta_{input}$ of each model is measured on its unperturbed magnitudes.}
         \label{fig4}
\end{figure}
\begin{figure}[!ht]
   \centering
 \includegraphics[width=9cm]{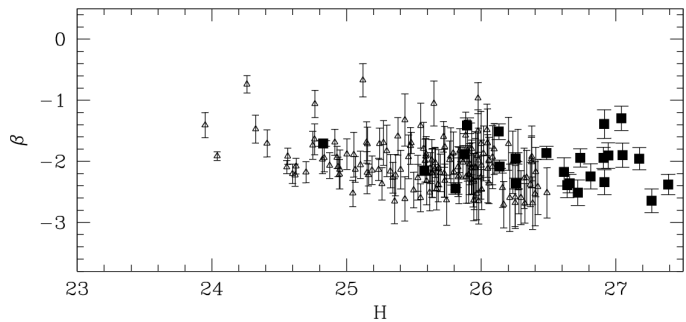}
   \caption{The UV slope $\beta$ of the ERS (open triangles) and HUDF (filled squares) B-dropout objects as a function of their observed H band magnitude. Errorbars are given by the uncertainty on the linear fit on their observed I, Z, Y, J AB magnitudes.}
         \label{fig5}
\end{figure}
 We adopt the common power-law approximation for the UV spectral range $F_{\lambda}\propto \lambda^{\beta}$, and we estimate the slope $\beta$ by fitting a linear relation through the observed I, Z, Y, J AB magnitudes of the objects:

\begin{equation}\label{betaformula}
 M_{i}=-2.5\cdot(\beta+2.0)\cdot log(\lambda_{i})+c
\end{equation} 

where $M_{i}$ is the magnitude in the i-th filter at effective wavelength $\lambda_{i}$. The filters we use span the  rest-frame wavelength range $\lambda\simeq~1500-2500$\AA~at $z\sim4$, providing a $\beta$ estimate which is consistent with the original definition by \citet{Calzetti1994}, and that is independent on the exact redshift of the sources, which only affects the intercept $c$ in Eq.~\ref{betaformula}. We underline that such estimate of $\beta$ is unbiased with respect to the colour selection being based on a different set of filters. We present in Fig.~\ref{fig3} four examples taken from the ERS sample, along with a comparison between the UV slopes determined by this method and the relevant best-fit SEDs.

The large wavelength baseline we adopt gives $\beta$ estimates that are significantly more stable than estimates based on the $I-Z$ colour alone \citep[e.g.][]{Bouwens2009a}: as an example, we show in Fig.~\ref{fig4} a comparison of the $\beta$ uncertainty distribution of the two methods for H$\sim$27 objects from the HUDF mock catalogues presented in Sect.~\ref{simul}. 

\begin{figure}
   \centering
 \includegraphics[width=9cm]{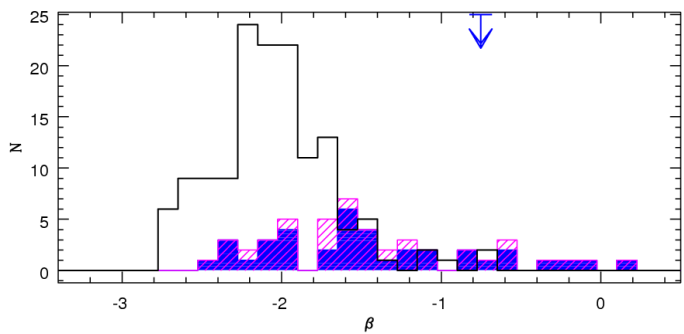}    
   \caption{Distribution of the UV slope values from the ERS sample (black hollow histogram). The magenta dashed histogram represent the objects that are not included in our sample but are selected through a standard (V-Z) colour-cut: most of them have $z_{phot}<3.5$ (blue filled histogram). The blue arrow marks the position of a $z_{spec}=3.11$ object included in this last sample.}
         \label{fig6}
\end{figure}

Fig.~\ref{fig5} shows the resulting UV slope $\beta$ for the objects in our sample as a function of their H band magnitude: most of the galaxies have blue UV slopes ($\beta \sim -2$), apart from a small number of red LBGs mainly populating the brightest end of the diagram.
We explore how the distribution of $\beta$ would change when adding to our sample objects that are included by a standard B-dropout selection diagram, but that are not selected when their (V-H) colour is considered. We perform this test on the ERS to explore also their photometric redshift distribution. As shown in Fig.~\ref{fig6}, these candidates (magenta dashed histogram) are on average redder than those already present in our sample, with 38 out of a total of 47 having a best-fit photometric redshift $z_{phot}<3.5$ (blue filled histogram). Moreover, the only spectroscopically confirmed object at $z_{spec}=3.11$ that we exclude through the (V-H) criterion has a very red UV slope with $\beta=-0.75$ (blue arrow in Fig.~\ref{fig6}). Notably, all the reddest objects ($\beta\sim0$) of the (V-Z) selected sample can be considered interlopers on the basis of their photo-z. These results are consistent with the analysis of LBG models presented in Sect.~\ref{selection}, showing that the usual (B-V)-(V-Z) cut preferentially selects objects with large E(B-V) at $z\sim3-3.5$.

A more detailed analysis of the LBG population can be given by looking at how UV slopes change as a function of the rest-frame UV magnitude: to this aim, a linear fit of the UV SED gives us a further advantage, namely we can interpolate the slope of each object to estimate the rest-frame magnitude at 1600\AA. To provide an estimate for $M_{1600}$, we consider all objects as located at the same redshift, which for simplicity we assume to be $z=4$: the effects of this assumption are self-consistenly taken into account in the statistical analysis discussed in the following sections.

\section{A statistical analysis of the UV slope distribution}\label{analysis}

\subsection{Uncertainties and systematics in $\beta$ and $M_{1600}$ estimates}\label{simul}

We first start with a set of Monte Carlo simulations reproducing the overall properties of our observations. 
We use the CB07 library to produce a large set of simulated galaxies at $3.0<z<5.0$ with expected magnitudes
in the same filter set as in the ERS and HUDF observations. We consider models with constant star-formation histories and within the following range of physical parameters: $Z/Z_{\odot}=0.02,0.2,1.0$, $0.01\leq E(B-V) \leq1.0$, $0.01 \leq Age \leq1$Gyr. We assume a \citet{Salpeter1955} IMF and a \citet{Calzetti2000} attenuation law, while the transmission of the IGM is treated according to \citet{Fan2006}.  We exploit this library to generate mock catalogues of $5\times10^5$ objects for each field by perturbing magnitudes in all the bands to match the relevant depths in the observed ones. Our simulations are designed to reproduce both the average and the scatter of the S/N distribution as a function of magnitude to provide the closest match between observed and simulated data.
The simulated catalogues are then analysed just as the real ones to determine the observational effects on galaxies of different magnitudes and spectral slopes.

Our Monte Carlo approach, while being much less time consuming than two-dimensional imaging simulations, allows us to estimate the effect of magnitude and colour uncertainties in the dropout selection and UV slope analysis. A further advantage of such a purely empirical approach is that it is free of assumptions on the typical size and size-luminosity relation of the LBGs, which, as shown in \citet{Grazian2011} are a critical factor to estimate the completeness correction for faint objects. We anyway randomly select a subsample of 20000 model LBGs from the same CB07 sample and perform a simple two-dimensional simulation to compare its results to those obtained through our empircal method.  We adopt as morphology template the stacking of ten bright candidates from the ERS field, and we perform various runs inserting 200 objects in the images each time after masking the regions where real objects were detected. The inserted objects are then recovered and analysed with the same procedure used for the real catalogues. A comparison between simulated objects obtained in this way and the mock catalogues discussed above shows a good consistency, therefore we will base our analysis on the latter ones that allow us to perform larger and less time-consuming simulations.   

\begin{figure}[!ht]
   \centering
\includegraphics[width=10cm]{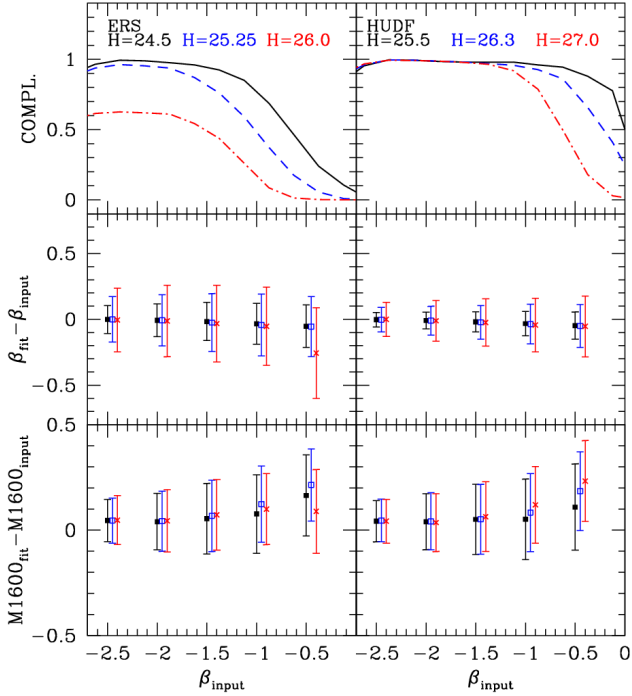}
   \caption{From top to bottom: detection completeness and difference between output and input $\beta$ and $M_{1600}$ from our ERS (left) and HUDF (right) mock catalogues. All quantities are computed for objects in three observed magnitude bins centred at: H=24.5 (ERS) and H=25.5 (HUDF) (black continuous lines and filled squares); H=25.25 (ERS) and H=26.3 (HUDF) (dashed blue lines and open squares); H=26.0 (ERS) and H=27.0 (HUDF)  (red dot-dashed lines and crosses).}
         \label{fig7}
\end{figure}
In the uppermost panel of Fig.~\ref{fig7}, we show for both fields the detection completeness at different observed fluxes of galaxies with a given spectral slope, as indicated by the $\beta_{input}$ of the model measured on its real unperturbed magnitudes. For the ERS sample the completeness curves remain approximately flat for $\beta_{input}\lesssim -1.5$ and rapidly decrease at $\beta_{input}\gtrsim -1$, especially for faint objects. The completeness in the HUDF is higher because of the deeper optical images: it remains close to one up to $\beta_{input}\lesssim -0.5$ for bright objects, while fainter ones are still efficiently selected as long as $\beta_{input}\lesssim -1.0$. The combination of the two fields thus allows us to sample a wide range of UV slopes and magnitudes.

In Fig.~\ref{fig7}, we also compare input and measured values of $M_{1600}$ and $\beta$ for galaxies taken from our simulations. In particular, we explore the dependence of uncertainties and systematics on the spectral shape of the galaxy, as indicated by $\beta_{input}$, and on the observed magnitude. The latter test is of particular importance since we are analysing the colour distribution of a flux-limited sample. 
The uncertainties on both $M_{1600}$ and $\beta$ are relatively small, with $\sigma_{M1600}\simeq0.1-0.2$ and $\sigma_{\beta}\simeq0.1-0.3$. However, they become larger for faint and red objects in the sample with respect to blue/luminous ones. The measured UV slope $\beta_{fit}$ does not show any large systematic offset as long as $\beta_{input}<-1$, a small trend towards underestimated values appears for the faintest red objects in the ERS sample. This is due to the shallower optical data in that field. The estimated UV rest-frame magnitude $M_{1600}$ shows a negligible offset of $\Delta\sim0.05-0.1~mags$ with respect to the real value for $\beta_{input}<-1$ with redder objects being underestimated by $\Delta\sim0.15-0.2~mags$. 
The conservative selection S/N(H)$>10$ helps keeping the uncertainties small: a similar test for objects of S/N(H)$\sim5-10$ shows that the uncertainty on the UV slope is as large as 0.6 for faint galaxies.

\subsection{Reconstructing the UV slope distribution: the method}
The tests discussed in Sect.~\ref{simul} show that the completeness and the accuracy on both $\beta$ and UV rest-frame magnitude estimates depend on the observed magnitude as well as on the colour of the galaxies. This implies that any \textit{measured} UV slope distributions at different luminosites arise from the underlying \textit{intrinsic} one after being convolved by a combination of observational effects. 
To estimate the dependence of $\beta$ on UV magnitude it is therefore not sufficient to apply simple corrections on the observed distributions but it is necessary to adopt a more comprehensive approach. Such an approach has to be based on assumptions on the properties of the intrinsic UV slope distribution, likewise any ``de-convolution'' procedure.
\begin{figure*}[!ht]
   \centering
   \includegraphics[width=18cm]{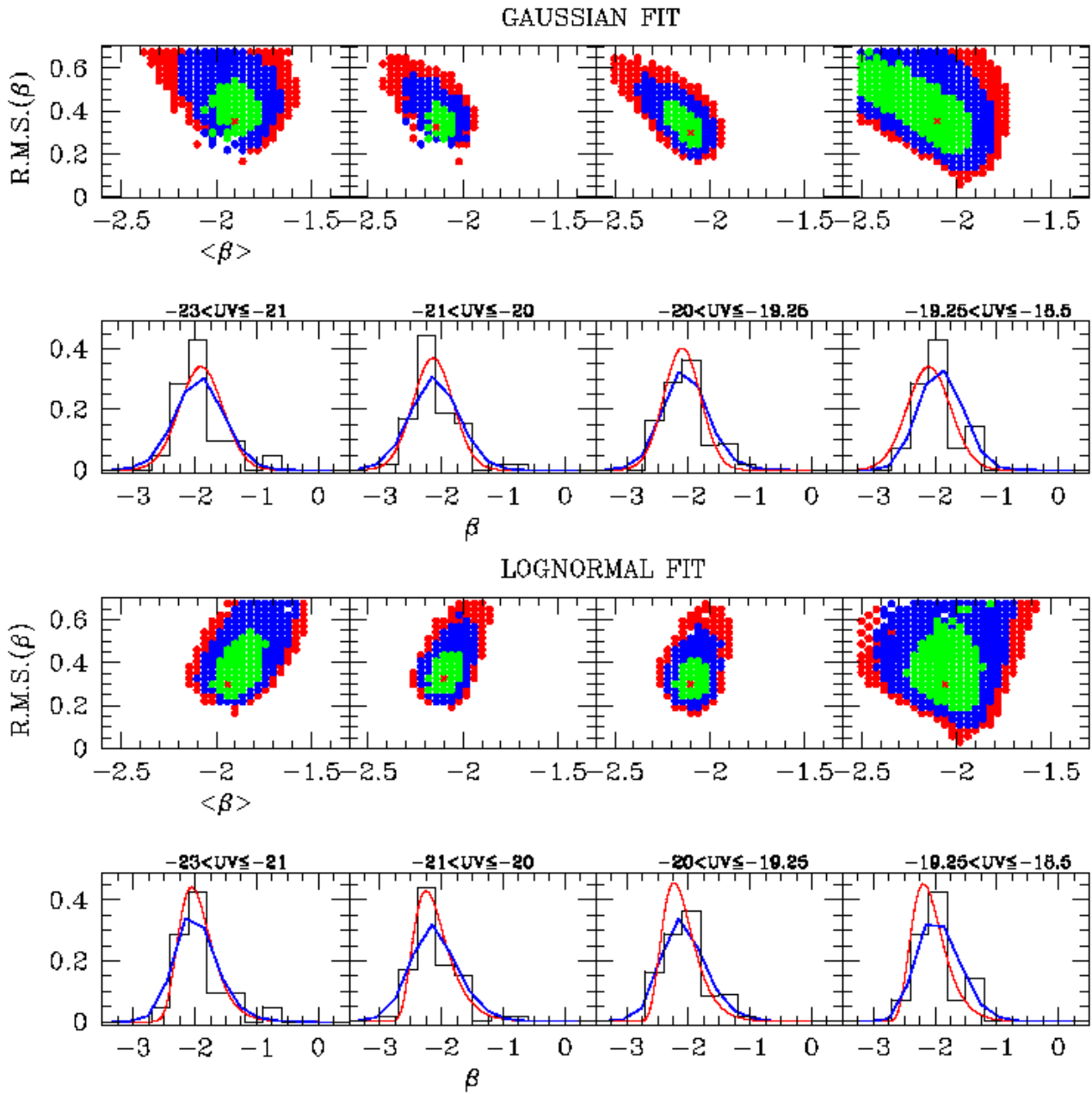}
   \caption{Maximum likelihood contours ($68\%$, $95\%$ and $99\%$ c.l. in green, blue, and red respectively) for the Gaussian and lognormal fit at different magnitudes. The histograms are the observed data. The red continuous line in each bin is the relevant best-fit distribution. The blue line show how it appears after being ``convolved'' by  observational effects.}
         \label{fig8}
\end{figure*}
We apply here a maximum likelihood technique which is analogous to the one used to estimate the galaxy Luminosity Function (LF) where a Schechter LF is assumed and its best-fit parameters are obtained by comparing simulated number counts to the observed ones.
We start from the assumption that the probability distribution function  of UV slopes in a given magnitude range, PDF($\beta$), follows a given functional form, whose average and standard deviation are to be determined. For any given set of parameters, and for each magnitude bin, we extract from our CB07 libraries 50000 objects reproducing the given distribution and perturb them with noise as discussed in Sect.~\ref{simul}, to introduce the effect of observational uncertainties and biases. Finally, we compare the output distribution of $\beta$ values to the observed one in each bin through a maximum likelihood estimator $\cal{L}$ \citep[e.g.][]{Bouwens2008,Castellano2010} and minimize $\Delta\chi^2=-2.0\cdot
ln(\cal{L})$.
Unfortunately, while a Schechter form in the case of the luminosity function is suggested by theoretical arguments and has been verified through many samples at low and high redshift, the present work constitutes the first attempt of a parametrical reconstruction of the UV slope distribution PDF($\beta$). We are thus forced to start from reasonable hypothesis on the analytic shape of PDF($\beta$), as suggested by our data and by previous non-parametric analysis.
We adopt two different analytic PDF($\beta$) as working hypothesis. We first assume a Gaussian distribution for the $\beta$ values. We note that such a distribution has already been suggested by other authors in the past \citep[e.g.][]{Bouwens2009a}. 
As an alternative, we assume a log-normal distribution PDF($\beta$) with $\beta=-2.5$ as lowest, asymptotic value. This is motivated by the asymmetric distribution observed in the brightest magnitude bins (Fig.~\ref{fig8}) and corresponds to the hypothesis that the high-z galaxy population is made of a bulk of low-extincted star-forming LBGs along with a number of rare objects at increasing E(B-V) values. 

We change the average and dispersion of both input distributions across a 25x25 grid in the range $\langle\beta\rangle=-2.5,-1.5 $ and r.m.s.($\beta$)=0.03,0.7.

\begin{figure*}[!ht]
   \centering
   \includegraphics[width=16cm]{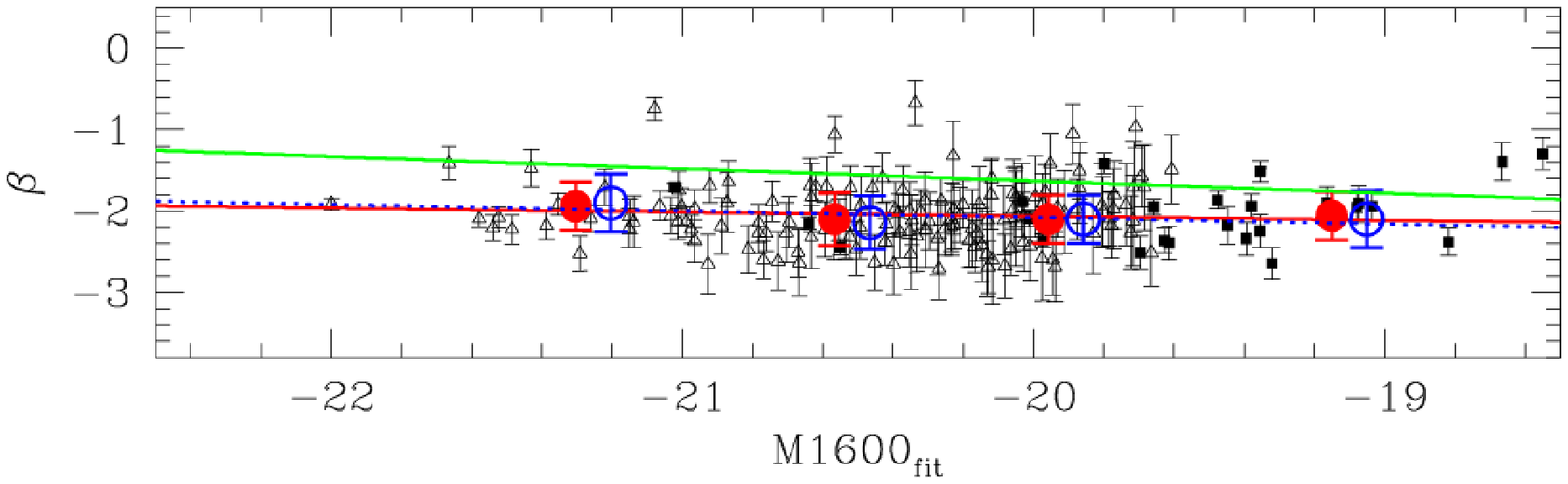}
   \caption{The $\beta-M_{1600}$ diagram for our B-dropout sample. Open triangles and filled squares represent individual objects in the  ERS and HUDF fields, respectively. Errorbars on the observed points are given by the uncertainty on the linear fit on their observed magnitudes. Blue and red circles and lines indicate the result of the statistical analysis described in Sect.~\ref{analysis}, which takes into account systematic effects. Blue open circles and errorbars indicate the average $\langle\beta\rangle$ and intrinsic r.m.s of the best-fit Gaussian distributions in each of the four bins: uncertainties on $\langle\beta\rangle$ are much lower and they are not visible in this scale. Red filled circles for the best-fit log-normal distributions. The Gaussians best-fit values are plotted at the average magnitude of the observed points in the relevant bins, log-normal ones with a small magnitude offset for clarity. The green continuous line is the $\beta-UV$ relation by \citet{Bouwens2009a}. We show for comparison a linear fit on the Gaussian and log-normal best-fit points (blue dashed and red continous lines): we underline however that our analysis indicates a significant difference only between the first and second magnitude bins.}
         \label{fig9}
\end{figure*}

\subsection{Results}
In order to analyse the relation between magnitude and UV slope, we divide the sample into four $M_{1600}$ bins (Fig.~\ref{fig8}). This choice is somewhat arbitrary and is motivated by the competing needs of having magnitude bins as narrow as possible, but with a sufficient number of objects allowing the statistical analysis described in the following. In practice, given the different magnitude ranges probed by the two fields, the ERS objects will be used to constrain the first and second bin, the third one will be analysed by combining data from both fields, while HUDF objects  will be used for the faintest magnitude bin. 

The resulting best-fit distributions (Table~\ref{bestfits}) indicate that most of the B-dropouts have blue UV slopes: in  particular, at $L<L^*$ galaxies tend to be very blue regardless of their $M_{1600}$, while the brightest objects only (first bin) are, on average, slightly redder than faint ones. Indeed, the best-fit Gaussians  are remarkably similar in the three magnitude bins at $M_{1600}>-21.0$ with an average of $\langle\beta\rangle\simeq -2.1$ and an intrinsic r.m.s. $\simeq 0.3$, while brighter galaxies have a best-fit Gaussian with $\langle\beta\rangle\simeq -1.9$, but similar dispersion. Given the low uncertainties, this small difference between bright and faint galaxies is significant at $\sim2\sigma$.
The log-normal fit yields similar results, with bright galaxies (first bin) redder than $M_{1600}>-21.0$ objects, the only difference being that in this case, given the asymmetric shape of the distribution, the bulk of the objects would have $\beta\sim-2.0$ in the first bin and $\beta\sim-2.2$ at $M_{1600}\gtrsim-21.0$. We verified that these results are robust against small changes in the colour selection criteria: objects selected with the redder boundary in the  (B-V) vs. (V-H) diagram discussed in Sect.~\ref{samplesubsect} yields best-fit values within the 68\% c.l. region given by our ``fiducial'' sample.

The binned data and best-fit distributions along with the maximum likelihood contours for the parameters are shown in Fig.~\ref{fig8}: the need for an approach like the one carried out here can be appreciated by looking at how the input best-fit distributions in the various magnitude bins (red curves in Fig.~\ref{fig8}) are modified when they are self-consistently ``convolved'' with observational effects through simulations (blue curves). 

Finally, Fig.~\ref{fig9} shows the distribution of observed points in the $\beta-M_{1600}$ diagram along with the average and intrinsic dispersion of the best-fit Gaussian and lognormal distribution.

While our tests consistently indicate that most of the B-dropouts have blue UV slopes, it does not allow us to determine which is the functional form that best reproduces the data.  To this aim, we compared the observed UV slope distributions to the best-fit Gaussian and log-normal PDF($\beta$) in each bin through a Kolmogorov-Smirnov test. We find a significant difference only in the first bin where the Gaussian fit yields a low KS probability ($<$50\%) because of the presence of a tail with few red objects at $\beta\simeq-1.0$. In the remaining bins the KS test does not allow us to discriminate between the two distributions.
It is interesting to note that a constant distribution of $\beta$ values within each bin is completely excluded by our maximum likelihood test at all magnitudes. However, we cannot exclude that the distribution is bimodal, with a ``red'' peak at values of $\beta$ larger than accessible to Lyman Break selections, or that it is a more complicated one, as it would be the case of Gaussians with non-standard kurtosis and/or skewness. While the first case can only be probed by combining LBG selected samples with far-IR and submillimetre selected ones \citep[e.g.][]{Wardlow2011}, the latter would require fitting more parameters than it is meaningful to do with the presently available data.

The physical interpretation of these results is not straightforward, since several factors determine the UV slope: dust reddening, varying ages, metallicity, IMF, and star formation histories \citep[e.g.][]{Bouwens2009a,Wilkins2011}. However, the effect of dust is by far the most relevant and with the presently available data it is not possible to constrain how much other properties contribute to the variation of the UV slope distribution with magnitude, or if they only affect the intrinsic dispersion of the PDF($\beta$) distributions. For this reason, and to provide comparison with other works, in the following we will interpret our results by directly converting the UV slope into dust extinction.

In this context, our results imply that the average extinction at 1600\AA~\citep[following][]{Meurer1999} is $A_{1600}\simeq0.65$ in the first bin and  $A_{1600}\simeq0.25$ in the faintest bins. Deviations from the \citet{Meurer1999} relation have been found in z$\gtrsim$4 QSOs by \citet{Gallerani2010}: their MEC attenuation curve would imply a higher attenuation, $A_{1600}\simeq 1.4$ and $A_{1600}\simeq 0.5$ for bright and faint galaxies respectively.

\begin{table}
\caption{UV slope distributions: best-fit parameters and 68\% c.l. uncertainties.}
\label{bestfits}
\centering
\begin{tabular}{ccc}
& GAUSSIAN&\\

\end{tabular}

\begin{tabular}{ccc}
\hline
UV magnitude & $\langle\beta\rangle$ & R.M.S.($\beta$) \\
\hline
$M_{1600}\leq$ -21.0& -1.90$^{+0.12}_{-0.08}$ & 0.35$_{-0.08}^{+0.19}$\\
 -21.0$<M_{1600}\leq$-20.0& -2.14$^{+0.08}_{-0.04}$ & 0.32$_{-0.05}^{+0.08}$\\
-20.0$<M_{1600}\leq$-19.25& -2.10$^{+0.04}_{-0.08}$ & 0.30$_{-0.05}^{+0.11}$\\
-19.25$<M_{1600}\leq$-18.5& -2.10$^{+0.06}_{-0.20}$ & 0.35$_{-0.13}^{+0.16}$\\
\hline
\end{tabular}

\begin{tabular}{c}
 LOGNORMAL\\

\end{tabular}

\begin{tabular}{cccc}
\hline
$M_{1600}$ & MAX. & $\langle\beta\rangle$ & R.M.S.($\beta$) \\
\hline
$M_{1600}\leq$ -21.0& -2.05 $^{+0.18}_{-0.17}$&-1.94$^{+0.16}_{-0.08}$ & 0.30$_{-0.05}^{+0.19}$\\
 -21.0$<M_{1600}\leq$-20.0& -2.25 $^{+0.10}_{-0.13}$& -2.10$^{+0.08}_{-0.08}$ & 0.32$_{-0.05}^{+0.11}$\\
-20.0$<M_{1600}\leq$-19.25& -2.23$^{+0.10}_{-0.12}$&-2.10$^{+0.08}_{-0.08}$ & 0.30$_{-0.05}^{+0.11}$\\
-19.25$<M_{1600}\leq$-18.5& -2.19$^{+0.19}_{-0.30}$ &-2.06$^{+0.16}_{-0.16}$ & 0.30$_{-0.11}^{+0.29}$\\
\hline
\end{tabular}
\end{table}

\subsection{Comparison with previous results}

Previous determinations of the UV extinction of $z\sim4$ LBGs have been provided by \citet{Ouchi2004},  \citet{Beckwith2006}, \citet{Overzier2008} \citet{Bouwens2009a} and \citet{Lee2011}.

\citet{Ouchi2004} found an average E(B-V)=$0.15\pm0.03$ for their $M<M^*$ B-dropouts and no dependence on the apparent (i.e. uncorrected) luminosity within the magnitude range they probe. Our estimate at similar magnitudes ($M_{1600}<-21.0$) is lower: when converting the relevant best-fit Gaussian PDF($\beta$) into a corresponding E(B-V) distribution following the equations in \citet{Calzetti2000}, we obtain an average $\langle E(B-V) \rangle = 0.05^{+0.03}_{-0.02}$. On the other hand, \citet{Lee2011} find a strong dependence of $\beta$ on UV magnitude in a large sample of $M<M^*$ $z\sim3.7$ galaxies from the NOAO Deep Wide Field Survey. Although they probe much brighter magnitudes than those available in our sample, a comparison can be made between their faintest bin centred at $M_{1700}=-21.43$ and the brightest galaxies from our ERS sample. They find $\beta=-1.78\pm0.28$ at these magnitudes which falls within the 68\% c.l. region given by our maximum likelihood test. 

\citet{Beckwith2006} find blue slopes for both bright and faint $z\sim$4 galaxies in the HUDF, suggesting that they are affected by little dust extinction. A quantitative comparison is not possible because \citet{Beckwith2006} do not explicitly compute the value of the UV slope in the different samples. However, their result is in agreement with our findings of blue UV slopes in all  $L\lesssim L_*$ magnitude bins, which is the range probed by the HUDF.

\citet{Overzier2008} provides an estimate of the UV slope of LBGs and Lyman Alpha Emitters (LAE) in a z$\sim$4.1 protocluster finding a global average $\langle \beta \rangle$=-1.95 which is in agreement with our results.

On the other hand, our results are markedly different from those obtained by \citet{Bouwens2009a}. They find redder $\beta$ averages and a steep correlation between $\beta$ and UV magnitude up to very faint luminosities: their relation is shown as a green line in Fig.~\ref{fig9}, for comparison we show as blue (red) lines a simple linear regression of our best-fit Gaussian (log-normal) average slope $\langle\beta\rangle$ as a function of $M_{1600}$.

These discrepancies can in part be explained by the differences
between various approaches, especially by the different colour
selections as discussed in Sect~\ref{sample}. Specifically, we explored whether the 
 greatest differences between our analysis and the \citet{Bouwens2009a} one can be responsible of the discrepancies:
we mimicked their approach by adopting the B-V, V-Z criterion (i.e. including in our sample the objects shown in Fig.~\ref{fig6}) and we
adopted the same definitions for $\beta$ and $M_{UV}$  based
on the I and Z observed magnitudes. 
With the inclusion of the additional red candidates,
at bright magnitudes  we obtain average UV slopes redder ($\beta\sim$-1.7) than in the case when
potential interlopers are excluded through their (V-H) colour.
However even with this inclusion, the values still are
not as red as those found by \citet{Bouwens2009a}\footnote[1]{We note that \citet{Bouwens2011} have recently presented a refined estimate of the average UV slopes where WFC3 IR information is also exploited, as in the present paper. Their findings show a good agreement with our estimates.}.

A completely different approach in the estimate of dust extinction in high redshift galaxies is provided by stacking analysis at radio wavelengths to determine their average uncorrected SFR.  In this respect, the analysis of $z\sim4$ LBGs by \citet{Ho2010}, based on VLA 1.4GHz observations, favors lower values for dust extinction than those inferred from past UV analysis, in agreement with our findings. However, a comparison between the two approaches is complicated by several factors and assumptions \citep[see discussion in][]{Carilli2008}.

\section{The dust corrected Star Formation Rate Density}\label{sfrd}
The parametric estimate of the UV slope distribution at different magnitudes gives us the remarkable opportunity of computing
the SFRD in a straightforward way through the following equation:
\begin{equation}\label{eqsfrd}
SFRD= \frac{1.0}{8\cdot 10^{27}}\int dL\int dA \cdot PDF(A,L) \cdot 10^{0.4 \cdot A} \cdot L \cdot \Phi(L)
\end{equation} 

where the constant factor is from \citet{Madau1998}, A=$A_{1600}$ such that $PDF(A,L)$ is univocally related to the $PDF(\beta,L)$ through the \citet{Meurer1999} equation, and $\Phi(L)$ is the UV luminosity function at 1600\AA. For the latter we adopt the best-fit Schechter parameters by \citet{Bouwens2007} for consistency with previous results.
In practice, we compute the integral in each luminosity bin by considering the relevant best-fit distributions PDF($\beta$), 
and extrapolating the last one up to $M_{1600}=-17.5$ (corresponding to $L=0.04L^*$).
When the Gaussian PDF($\beta$) are considered we obtain log(SFRD)=$-1.093 ^{+0.18}_{-0.18}~M_{\odot}/yr/Mpc^3$, taking into account both the 68\% c.l. uncertainty on each PDF($\beta$) and the uncertainty on the LF parameters. Using the best-fit lognormal PDF($\beta$) in Eq.~\ref{eqsfrd} we obtain a slightly higher value log(SFRD)=$-1.086 ^{+0.20}_{-0.18}~M_{\odot}/yr/Mpc^3$ because of the larger effective correction for dust extincion. Our SFRD estimates are $\sim$55\% lower than the one by \citet{Bouwens2009a} : this not a surprise since we find bluer $\beta$ at bright and intermediate magnitudes, hence we apply a lower correction for dust. Interestingly, our SFRD values are consistent with the estimate obtained by \citet{Thompson2006} on a photometric-redshift selected $z\sim$4 sample and through the use of SED-fitting to estimate extinction correction. This is in agreement with the discussion in Sect.~\ref{sample} and~\ref{slopes}~on the differences between our H-based selection and the usual optical-based one.

The largest uncertainties in the SFRD estimate are given by our lack of knowledge of the attenuation curve at these redshifts, and by the possible presence of highly-obscured star-forming objects, like the submillimetre galaxies (SMGs) well studied at lower redshifts \citep[e.g.][]{Chapman2005}. 
As an example, the MEC attenuation curve by \citet{Gallerani2010}, which is flatter than the Calzetti et al. one in the UV, yields a SFRD nearly three times larger: log(SFRD)$\simeq -0.65~M_{\odot}/yr/Mpc^3$. On the other hand, a significant contribution from $z\sim4$ SMGs \citep[e.g.][]{Daddi2009} would imply higher values of the SFRD than those measured from LBGs alone.

\section{Dust extinction as a function of intrinsic UV luminosity}\label{UVcorr}
\begin{figure*}[!ht]
   \centering
   \includegraphics[width=16cm]{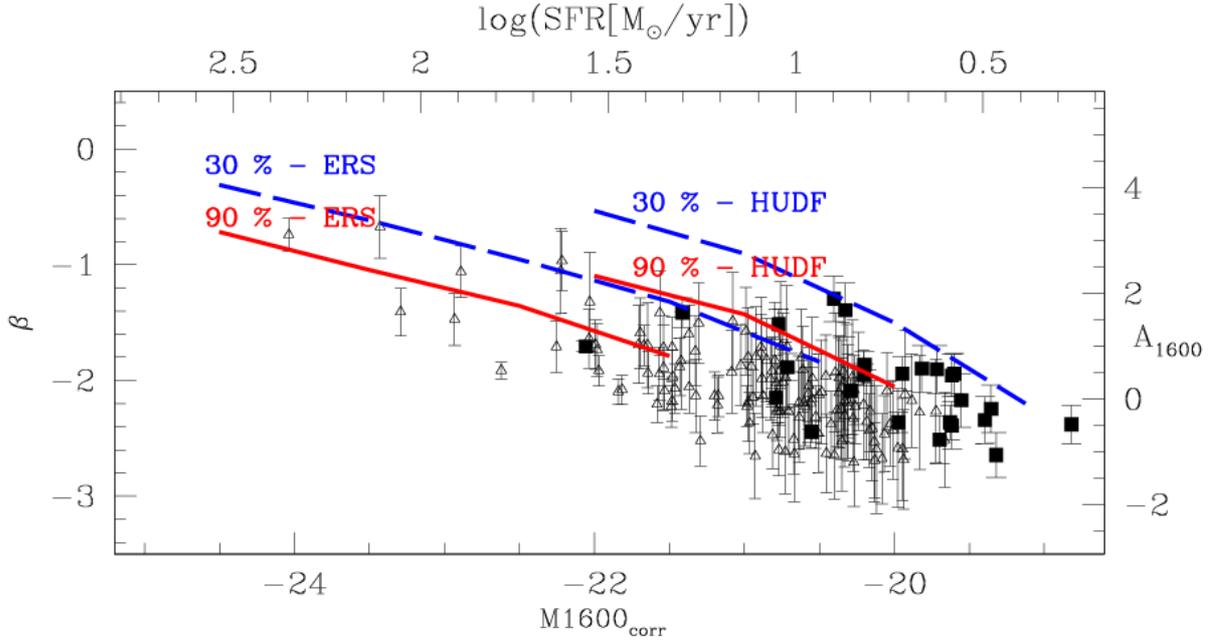}
   \caption{The relation between $\beta$ and the dust-corrected UV magnitude for ERS (open triangles) and HUDF objects (filled squares) in our sample. The upper and the right axes show the conversion of unextincted magnitude and UV slope in SFR \citep{Madau1998} and $A_{1600}$ \citep{Meurer1999} respectively. The red continuous lines and the blue dashed ones indicate the 90\% and 30\% selection completeness in the two fields. }
         \label{fig10}
\end{figure*}
Galaxy evolutionary models suggest that UV bright galaxies tend to be dustier than faint ones, 
as a natural outcome of the relation between SFR, mass and metallicity. 
Our findings of a redder UV slope distribution at bright magnitudes apparently confirm this prediction. However,  such analysis shows how galaxies of different UV slopes (which is a proxy of their dust content) are distributed as a function of their \textit{extincted} UV magnitude. The effect of extinction is thus present on both quantities and a clear characterization of how the dust content varies across the galaxy population is not possible: the usual $\beta$-UV magnitude relation, as presented in Sect.~\ref{analysis}, is mainly useful to estimate a correction to the observed SFRD, as discussed above, but do not offer a straightforward insight on the physical properties of Lyman break galaxies.

To provide a clearer view on this issue we show in Fig.~\ref{fig10} the relation between $\beta$ and the \textit{dust-corrected} UV magnitude $M_{1600corr}=M_{1600}-A_{1600}$. We also show the conversion of both in more grounded physical quantities: $A_{1600}$ \citep[following][]{Meurer1999} and SFR \citep{Madau1998} respectively. Naturally, degeneracies between extinction and other physical parameters can bias this direct conversion, and quantities are not completely independent, since the extinction correction is itself related to the $\beta$ of the single objects. Nonetheless, the simplicity of our analysis provides a more straightforward view on such observational degeneracies than in the case of common SED-fitting estimates.  We also plot in Fig.~\ref{fig10} the 30\% and 90\% detection completeness levels, as estimated from our simulations, for galaxies of the two fields. 

From this plot it appears evident that at large intrinsic luminosities ($M_{1600corr}\lesssim-23.0$, corresponding to SFR$\gtrsim80~M_{\odot}/yr$), only dusty star-forming galaxies are found. 
Once the incompleteness of our selection is taken into account we obtain for objects with SFR$>80~M_{\odot}/yr$  a surface density of 0.12~$arcmin^{-2}$ when $2<A_{1600}<3$ and an upper limit of 0.02~$arcmin^{-2}$ for less extincted galaxies.

Among galaxies with lower SFR, we detect many with a much lower amount of reddening. As a result  a trend of decreasing extinction as a function of corrected UV luminosity seems to be in place. However, this effect may be entirely a systematic effect due to the limits of our observations (as shown by  the lines in Fig.~\ref{fig10}) that do not allow us to ascertain whether this trend is real or it is due to selection effects even at moderately high luminosities. 

\section{Summary and Conclusions}\label{summary}
In this paper we have presented a new approach to the selection and to the analysis of the UV slope distribution of z$\sim$4 Lyman Break galaxies. Thanks to the deep WFC3 near infrared observations of the ERS and HUDF fields, combined with the existing HST-ACS optical data, we perform a B-dropout selection that is based on a (B-V) vs. (V-H) diagram on H-detected samples of both fields. Following the analysis of galaxy spectral libraries and of our own photometric redshift  catalogue on the ERS, we have shown that this selection allows us to efficiently exclude low redshift interlopers and dusty z$\gtrsim$3 galaxies that might be present in B-dropout samples selected through optical colours.
The sample we select on the basis of the newly defined colour criteria comprises 142 and 25 objects at S/N(H)$>$10 in the ERS and HUDF fields respectively. We estimate the UV slope of our B-dropout candidates through a linear fit on their I, Z, Y, J magnitudes, and we interpolate the same fit to provide an estimate of their UV rest-frame magnitude after assuming that all sources are located at z=4.

Despite the conservative selection and the greater accuracy in the estimate of $\beta$ provided by a linear fit, we show that uncertainties and systematics arise when measuring the colour distribution of such flux-limited samples. However, the effect of observational uncertainties can be effectively taken into account by adopting a parametric approach combined with detailed simulations, and the estimate of the typical UV slope at different magnitudes can be performed following hypothesis on the shape of the undelying intrinsic distribution.
We carry out a parametric maximum likelihood analysis by assuming that the distribution of $\beta$ is either Gaussian or lognormal. We allow the relevant parameters of the two distributions to vary across a wide range of values and find the best-fit values by comparing the simulated and observed counts through a maximum-likelihood estimator.

We find that $z\sim$4 LBGs are characterized by blue UV slopes, suggesting a low dust extinction. The best-fit Gaussians  are remarkably similar in all the magnitude bins at $M_{1600}>-21.0$ with an average of $\langle\beta\rangle\simeq -2.1$ and an intrinsic r.m.s. $\simeq 0.3$, while brighter galaxies have a best-fit Gaussian with slightly redder average ($\langle\beta\rangle\simeq -1.9$) but similar dispersion. A similar result is obtained when log-normal distributions are considered, the main difference being that in this case, given the asymmetric shape of the distribution, the bulk of the objects would have $\beta\sim-2.0$ in the first bin and $\beta\sim-2.2$ at $M_{1600}\gtrsim-21.0$. When interepreted in term of varying dust extinction, our results imply that z$\sim$4 LBGs have a low dust content, with  a typical $A_{1600}\simeq0.65$ at $L< L^*$ and  $A_{1600}\simeq0.25$ for faintest galaxies, following \citet{Meurer1999}.

Our results are in general agreement with the previous findings of \citet{Beckwith2006} and \citet{Overzier2008}, and are consistent within the uncertainties with the result on $L\sim L^*$ galaxies by \citet{Lee2011}. On the other hand our results differ from those obtained by \citet{Ouchi2004} for bright galaxies, and from \citet{Bouwens2009a} that find redder $\beta$ averages and a steeper correlation between $\beta$ and UV magnitude.  We find that these discrepancies are only in part due to the differences in the selection (IR-based vs. optically-based) and in the analysis of the B-dropout samples, since such red averages for the UV slope are not found in our sample even in the case when potential interlopers selected through their optical colours are included.

The low dust extinction we find points to a SFRD in the LBG population that is  significantly revised downward with respect to previous estimates.
If we compute the corrected SFRD from the UV LF by exploiting our parametric estimate (Eq.~\ref{eqsfrd}), we obtain log(SFRD)$\sim -1.09$, which is $\sim$55\% lower than the \citet{Bouwens2009a} estimate.
However, this result is based on the assumption of a Calzetti et al. attenuation law: when the MEC attenuation curve of \citet{Gallerani2010} is assumed we obtain a nearly three times higher SFRD at $z\sim$4. 

Finally, we discuss how the UV slope of z$\sim$4 galaxies changes as a function of the dust-corrected UV magnitude, i.e. as a function of the SFR.  Because of the difficulties in selecting highly extincted objects, we show that constraints can only be given on the number density of UV bright galaxies (SFR$\gtrsim80~M_{\odot}/yr$). We find that highly extincted galaxies with $2<A_{1600}<3$ are much more common (surface density of 0.12~$arcmin^{-2}$) than dust-free ones ($<0.02~arcmin^{-2}$). This implies that most galaxies with a high  SFR ($\gtrsim80~M_{\odot}/yr$) are highly extincted objects. Among galaxies with lower SFR, we detect many with a much lower amount of reddening, although  current observational limits prevent us from detecting those with high  exctinction, if they exist.

Although the now available WFC3 observations allow for a refined selection of z$\sim$4 samples, improvements are needed to further constrain the dust properties at high redshifts. Deeper spectroscopic follow-up observations are necessary to validate both IR and pure optical colour selections. Such samples will also give the opportunity of complementing the study of UV slope distributions with the analysis of the Ly$\alpha$ EW distribution which might also be related to the dust content of galaxies \cite[e.g.][]{Pentericci2007}.
Larger near IR data-samples, as those that will be available in the near future thanks to the CANDELS survey, will allow us to deepen this kind of analysis. While CANDELS/Wide fields will enable a comparison of the UV slope distribution of IR-selected LBGs to the past optical-based analysis of large L$>>L^*$ samples, the CANDELS/Deep observations will allow us to strengthen statistical constraints at faint/intermediate magnitudes.

Instead, a thorough characterization of the high-redshift population and a comprehensive analysis of the relation between SFR and dust content will require future instruments both in the optical and in the IR capable of observing highly extincted objects with SFR of few solar masses per year. 

\begin{acknowledgements}
We acknowledge the support of ASI-INAF in the framework of the program I/009/10/0 supporting Data Analysis in the field of Cosmology.

\end{acknowledgements}

\bibliographystyle{aa}

\end{document}